\documentclass[9pt,sigconf]{acmart}

\usepackage{graphicx}
\usepackage{textcomp}
\usepackage{xcolor}
\usepackage{subcaption}
\usepackage{svg}
\usepackage{float}
\usepackage{multicol}
\usepackage{array}
\usepackage{mwe}
\usepackage[noend]{algpseudocode}
\usepackage[ruled, lined, longend]{algorithm2e}
\usepackage{multirow}
\usepackage{arydshln}
\usepackage{adjustbox}
\usepackage[thinc]{esdiff}
\usepackage{booktabs} 
\usepackage{pifont}
\usepackage{hyperref}
\usepackage{todonotes}
\usepackage{enumitem}
\usepackage{tablefootnote}
\usepackage{threeparttable}
\usepackage{multirow}
\usepackage{amsmath}
\usepackage{siunitx}
\usepackage{graphicx}

\AtBeginDocument{%
  }

\setcopyright{none}

\begin{document}




\title{PD-Swap: Prefill–Decode Logic Swapping for End-to-End LLM Inference on Edge FPGAs via Dynamic Partial Reconfiguration}

\author{Yifan Zhang, Zhiheng Chen, Ye Qiao, and Sitao Huang}
\email{{yifanz58, zhihenc5, yeq6, sitaoh}@uci.edu}
\affiliation{%
  \institution{University of California, Irvine}
    \city{Irvine}
  \state{California}
  \country{USA}
}

\begin{abstract}
Aggressively quantized large language models (LLMs), such as BitNet-style 1.58-bit Transformers with ternary weights, make it feasible to deploy generative AI on low-power edge FPGAs. However, as prompts grow to tens of thousands of tokens, edge hardware performance drops sharply with sequence length due to quadratic prefill cost and rapidly increasing KV-cache bandwidth demands, making inference latency of longer context length a first-order system concern. Recent studies on LLMs expose a fundamental \emph{prefill--decode asymmetry}: prefill is compute-bound and dominated by dense matrix--matrix operations, whereas decoding is memory-bandwidth-bound and dominated by KV-cache traffic. A static accelerator must provision resources and a single dataflow for both regimes, leading to duplicated attention logic, underutilized fabric, and tight LUT/URAM limits that cap model size and usable context. We propose a prefill--decode disaggregated LLM accelerator, PD-Swap, that uses Dynamic Partial Reconfiguration (DPR) to \emph{time-multiplex} the attention module on edge FPGAs. The core table-lookup ternary matrix multiplication and weight-buffering engines remain static, while the attention subsystem is a reconfigurable partition with two phase-specialized architectures: a compute-heavy, token-parallel prefill engine and a bandwidth-optimized, KV-cache-centric decoding engine. A roofline-inspired model and design space exploration jointly optimize reconfigurable-region size, parallelism under reconfiguration and routability constraints, and reconfiguration latency is hidden by computation latency. PD-Swap achieves up to 27~tokens/s decoding throughput, outperforming prior state-of-the-art works by 1.3$\times$--2.1$\times$ (larger gains at longer context lengths), without extra area cost.
\end{abstract}

\maketitle

\section{Introduction}

Transformer-based large language models (LLMs) underpin many modern AI services, but their computation, memory, and bandwidth demands clash with the strict power and cost budgets of edge devices. Quantization is a key enabler for on-device LLM inference: BitNet-style 1.58-bit models show that ternary weights ($\{-1,0,+1\}$) can approach full-precision accuracy while drastically reducing model size and replacing multiplications with low-cost operations. Combined with the reconfigurability and fine-grained parallelism of FPGAs, such models offer a promising path toward privacy-preserving, low-latency LLM inference at the edge.


Recent works~\cite{moitra2025meadow,qiao2025tellme} have implemented end-to-end LLM accelerators with edge FPGAs, and they accelerate both prefill and autoregressive decoding on chip under tight power budgets and achieve competitive tokens/s compared to INT8 and FP16 designs. However, study on these end-to-end accelerators with both prefill and decode reveals a fundamental \emph{prefill-decode asymmetry}. Prefill processes the entire prompt in parallel and is dominated by matrix-matrix operations, making it compute-bound and constrained by LUT/URAM budget and timing closure. Decoding generates one token at a time, repeatedly accessing the KV cache and weights; its arithmetic intensity drops sharply and performance becomes dominated by DDR bandwidth, which is quickly saturated even with 4-bit quantization. A static edge accelerator must therefore provision hardware and a single dataflow for both regimes, duplicating attention logic, control, and buffering and limiting model size, frequency, and usable context length.

Modern FPGAs, including AMD Zynq and Versal families, support \emph{Dynamic Function Exchange} (DFX), a vendor-integrated form of dynamic and partial reconfiguration that allows part of the fabric to be reprogrammed while the rest continues to operate. In the DFX flow, the design is split into a static region and one or more reconfigurable partitions (RPs) that can host multiple reconfigurable modules (RMs) loaded via partial bitstreams. For modest RP sizes, reconfiguration can complete in milliseconds. 
Recent works have explored DPR-based FPGA accelerators for CNNs and small-scale neural networks on edge devices~\cite{dpr_fpl21,dpr_hpec19,dpr_fpl19}. However, these designs mainly target vision workloads with static computation patterns, and do not address the highly asymmetric and dynamic compute and memory characteristics in autoregressive LLM inference.

The prefill-decode asymmetry in LLM inference is a natural fit for \emph{logic swapping} on edge FPGAs. In our design, the ternary table-lookup MatMul and weight-buffering engines, which are shared by both phases, reside in the static region, while the attention subsystem is implemented as a reconfigurable partition with two RMs: a compute-heavy, highly parallel prefill engine and a bandwidth-optimized decoding engine with KV-cache-centric dataflow and tailored AXI/DDR scheduling. Prefill runs with the prefill RM loaded; once the prompt is processed, the decoding RM is loaded, and the system switches to generation mode. The reconfigurable region is sized so that partial bitstreams can be loaded within a small fraction of end-to-end latency, and reconfiguration is overlapped with ongoing computation in the static region to hide its cost.

We develop a roofline-inspired analytic model and design space exploration (DSE) flow that jointly optimizes prefill compute time and decoding throughput. Given a target device and LLM configuration, the DSE selects the size of the reconfigurable partition, the degree of parallelism in each attention RM, and the allocate resources under area, routability, and frequency constraints.
This paper makes the following contributions:
\begin{itemize}[nosep, leftmargin=*]
    \item We propose a prefill-decode disaggregated LLM accelerator for edge FPGAs that \emph{time-multiplexes} the attention hardware module via dynamic partial reconfiguration, achieving optimized latency for both prefill stage and decode stage without consuming extra hardware resources compared to static accelerators.
    \item We present a roofline-inspired analytic model and design space exploration (DSE) flow that jointly optimize prefill and decode stage mapping under tight area constraints on edge devices.
    \item We design a latency-overlapped runtime reconfiguration mechanism that exploits the structure of LLM inference to reduce reconfiguration overhead to a negligible level.
    \item We prototype our design on AMD Kria KV260 and it outperforms state-of-the-art works: up to 27~tokens/s generation throughput and improved time-to-first-token, achieving $1.3\times$--$2.1\times$ higher decoding throughput than state-of-the-art static accelerators.
\end{itemize}

\begin{figure}
    \centering
    \includegraphics[width=1\linewidth]{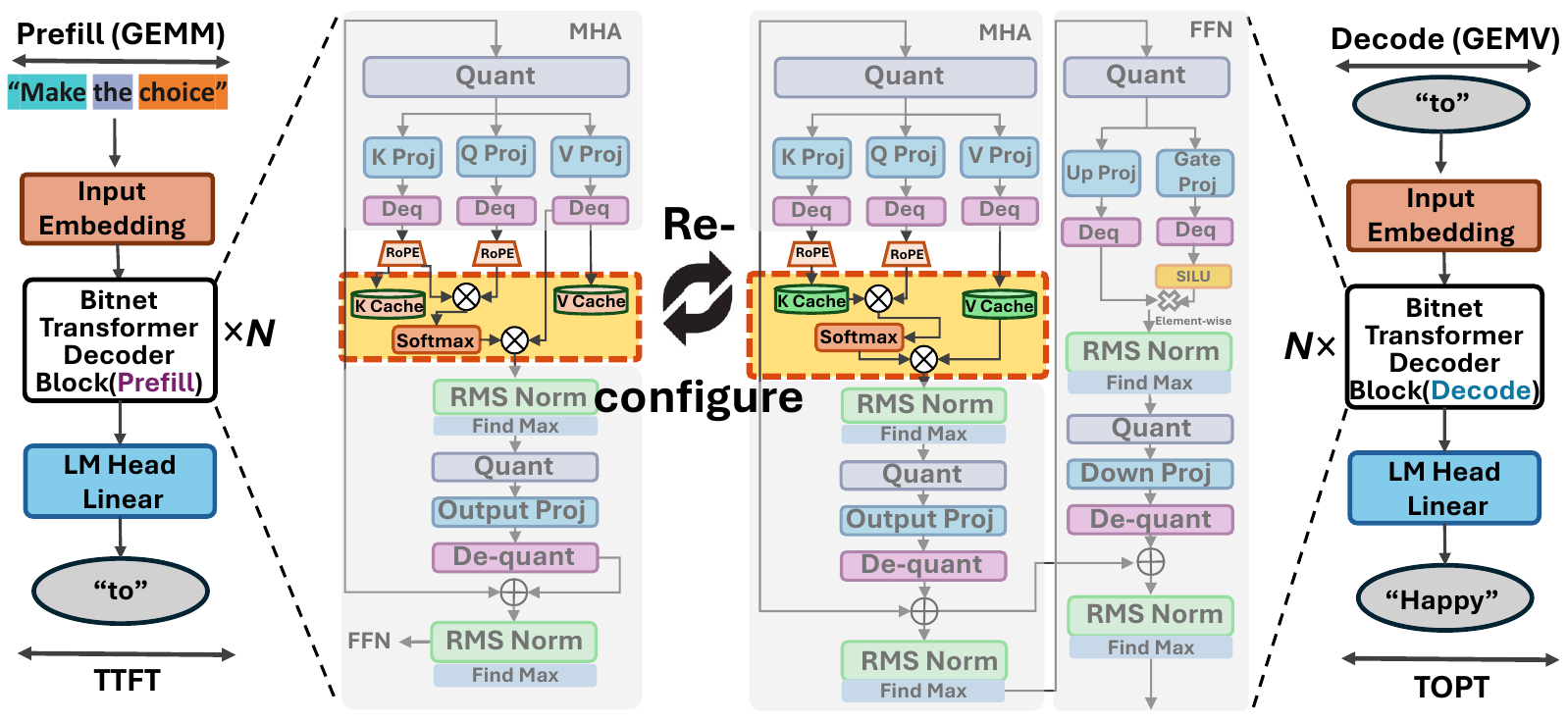}
    \caption{PD-Swap Prefill/Decode Stages in LLM Architecture}
    \label{fig:model_arch}
\end{figure}

\section{Background and Related Work}

\subsection{LLM Prefill--Decode Asymmetry}

Transformer LLMs perform autoregressive inference in two phases: \emph{prefill}, which processes a prompt of length $N$ in parallel, and \emph{decoding}, which generates one token at a time using the accumulated KV cache. Prefill is dominated by matrix--matrix operations in attention and feedforward layers and is typically compute bound, while decoding reduces these to matrix--vector and vector--vector kernels but repeatedly streams weights and KV cache from memory, making it bandwidth bound. Analytic studies of FPGA-based spatial LLM accelerators formalize this behavior and show that FPGAs lag GPUs in compute-intensive prefill but can be competitive or superior in memory-bound decoding when KV-cache traffic dominates the roofline~\cite{chen2023spatial}. Embedded implementations of 4-bit LLaMA on KV260 further confirm that decoding quickly saturates DDR bandwidth, whereas prefill is primarily limited by on-chip parallelism and buffering~\cite{li2025pushing,li2025hummingbird}. These results indicate that the optimal hardware configuration and dataflow for prefill and decoding are fundamentally different, yet most designs rely on a single static architecture that must compromise between them.

\subsection{Ternary LLMs and Table-Lookup Matmul}

BitNet-style 1.58-bit LLMs quantize Transformer weights to ternary values $\{-1,0,+1\}$ and can match FP16/BF16 accuracy while substantially reducing model size and compute cost~\cite{ma2024bitnet}. On FPGAs, ternary weights enable table-lookup matmul (TLMM), which packs ternary weights into compact indices and precomputes partial sums in LUT/URAM-resident tables so that runtime matmul becomes index--lookup--accumulate~\cite{qiao2025tellmev2}. Recent TLMM-based accelerators on AMD Kria KV260 demonstrate end-to-end 1.58-bit LLM inference within a 5--7~W power envelope and competitive tokens/s relative to INT8 designs, but also show that DSP usage is largely eliminated and LUT/BRAM/URAM become the primary bottlenecks~\cite{qiao2025tellme,qiao2025tellmev2,qiao2025cobra}. Prefill latency in these designs is dominated by TLMM throughput and attention compute, while decoding is dominated by DDR accesses for TLMM weights and KV cache. Because prefill and decoding attention pipelines are instantiated statically, a large amount of logic and buffering is duplicated, leaving fewer resources for TLMM and on-chip weight and tightening timing closure on small devices.

\subsection{FPGA-based LLM Accelerators on Edge}

A growing set of works target LLMs on edge-class FPGAs. SECDA-LLM introduces a design platform and a block-floating-point matmul engine for TinyLLaMA on PYNQ-Z1, demonstrating substantial speedups over a CPU baseline but focusing on a single kernel rather than full LLM pipelines~\cite{haris2024secda}. MEADOW proposes a token-parallel, head-sequential dataflow and lossless weight packing for INT8 models on ZCU102, reducing off-chip accesses and improving prefill and decode latency versus GEMM-based baselines, but without phase-specialized hardware~\cite{moitra2025meadow}. On larger boards, EdgeLLM uses approximate FP16$\times$INT4/FP16$\times$FP16 engines in a CPU--FPGA heterogeneous system and achieves better throughput and energy efficiency than an A100 GPU for ChatGLM2-6B, leveraging HBM and abundant resources not available on embedded SoCs~\cite{huang2024edgellm}. Spatial accelerators on Alveo U280 further show that FPGAs can be competitive with GPUs for LLMs but remain bandwidth limited in decoding for long contexts~\cite{chen2023spatial}. Embedded accelerators for LLaMA2/3 on KV260/ZCU104 aggressively optimize memory layout and AXI arbitration to reach multi-token/s decoding at high DDR utilization, yet their architectures are static and tuned primarily for decoding, with prefill either simplified or offloaded~\cite{li2025pushing,li2025hummingbird}. Ternary TLMM-based accelerators complement these INT4/INT8 designs by further compressing weights and exploiting LUT/URAM, but exacerbate contention between matmul, attention, and normalization logic~\cite{qiao2025tellmev2}. Across these works, prefill and decode are still treated as a single monolithic workload mapped onto a fixed microarchitecture.

\subsection{Dynamic and Partial Reconfiguration}

Dynamic and partial reconfiguration (DPR) allows updating a subset of the FPGA fabric at runtime while the rest of the design continues to operate. Vipin et al. survey DPR architectures, configuration interfaces, and use cases, and report that partial bitstreams for modest regions can be loaded in tens of milliseconds on Zynq-class devices when configuration bandwidth is fully utilized~\cite{vipin2018survey}. DPR has been applied to multi-standard radios, adaptive filters, cryptographic cores, and multi-tenant overlays, where accelerators or algorithm variants are swapped at coarse time scales. More recently, high-level flows ~\cite{xiao2022hipr,xiao2022pld, pr_esp} integrate DPR into system-level design.
PLD uses PR-style partitioning to enable separate compilation and linking of accelerators for 3--10$\times$ faster incremental builds~\cite{xiao2022pld}, while HiPR introduces C-level pragmas that mark functions as reconfigurable and automatically generates static and reconfigurable regions and partial bitstreams~\cite{xiao2022hipr}. These frameworks mainly exploit DPR for faster development and modular deployment; they do not target runtime adaptation across distinct phases of a single workload.

In the LLM domain, existing FPGA accelerators are effectively static: logic and routing remain fixed throughout inference, and adaptation to prefill or decoding is handled via scheduling and software-level control. To the best of our knowledge, no prior work combines LLM acceleration with time-multiplex attention logic between prefill and decoding on embedded FPGAs. This gap motivates our prefill--decode disaggregation architecture with DPR-based logic swapping and an analytic model that explicitly accounts for both reconfiguration and inference latency.

\section{PD-Swap Design Methodology}
\subsection{Overview}
The goal of PD-Swap is to boost the performance of end-to-end LLM inference on the resource-constrained edge FPGA. 
We adopt BitNet-1.58 as the baseline LLM architecture for our system. BitNet employs mixed-precision arithmetic and ultra-low-bit ternary weights, substantially reducing memory bandwidth requirements and enabling LLM deployment on resource-constrained edge FPGA platforms.
Meanwhile, the heterogeneity of the various modules in the BitNet model necessitates dedicated data flow optimization. 

Figure~\ref{fig:model_arch} illustrates PD-swap in the BitNet model architecture. The major workload in BitNet Model inference includes attention computation and RMSNorm (both fp16), linear computation with quantization/dequantization (ternary weights, INT8 activation). Other element-wise operations include RoPE and SwiGLU.

\subsection{PD-Swap Architecture Design}

\begin{figure}
    \centering
    \includegraphics[width=1\linewidth]{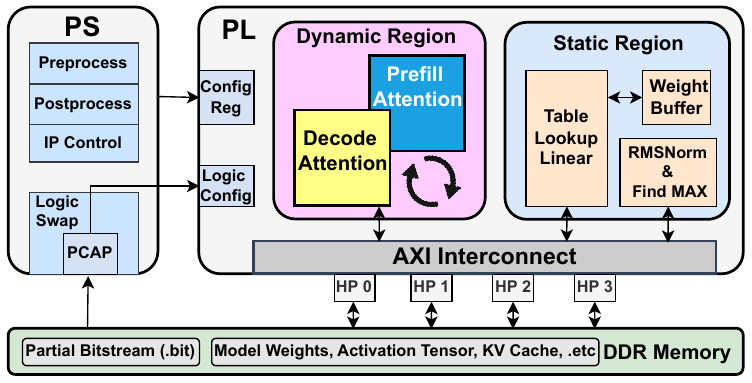}
    \caption{PD-Swap System Design Overview}\vspace{-2em}
    \label{fig:system_diagram}
\end{figure}

Besides the heterogeneous workload in Bitnet model, our system design is also motivated by the observation that the prefill and decoding stages of LLM inference exhibit fundamentally different computational characteristics. The prefill stage processes the entire input prompt in parallel and is dominated by compute-bound attention operations whose complexity scales quadratically with the sequence length. In contrast, the decoding stage handles only one token per iteration, making it inherently memory-bound; the projection layers remain constant in cost while the attention layers scale linearly with the accumulated sequence length. These distinct patterns indicate that a single, monolithic hardware design cannot simultaneously achieve optimal performance across both stages.
To exploit this asymmetry, we architect a runtime-reconfigurable inference system on edge FPGA that partitions the device into a static region and a dynamic region. Figure~\ref{fig:system_diagram} illustrates the overall system architecture. The static region hosts operators whose computation patterns remain stable across different phases of inference, including the projection layers, RMSNorm, and other element-wise opeartors. These operators benefit little from hardware specialization across stages, making them well-suited for persistent deployment. In contrast, the dynamic region is dedicated to the attention computation, whose optimal hardware structure differs significantly between prefill and decoding. 

\subsubsection{Runtime Reconfiguration Workflow}

Instead of reconfig the datapath through control signal, we directly swap the logic in the dynamic region to the optimal configuration, which is enabled by the DPR technique.
During design time, the FPGA fabric is partitioned into a static region and a dynamic region using pblocks, where the dynamic pblock forms an RP. Vivado then generates the partial bitstreams for PR and fixes RP pins to ensure compatibility. DPR allows the rest of the design to continue running during RP reconfiguration.
By leveraging DPR, our design replaces the attention accelerator at runtime. A compute-optimized prefill attention module is loaded during the prompt processing phase, while a memory-optimized decoding attention module is swapped in once token generation begins.
 A global inference controller on the Processing System (PS) orchestrates stage transitions by monitoring the model execution flow and initiating partial reconfiguration when entering or leaving the prefill stage. The dynamic region is sized to accommodate the largest attention variant, while the static region provides a stable interface for memory transactions, model parameter access, and system-level buffering. 
PS  streams the partial bitstream into the FPGA via the Processor Configuration Access Port (PCAP), overwriting only the dynamic region with the new hardware. Once reconfiguration is done, the PS re-enables the interfaces and triggers the decode attention kernel.

Moreover, a partial bitstream is significantly smaller than a full device bitstream, enabling reconfiguration within milliseconds through the PS-to-Programming Logic (PL) configuration channel. This workflow enables efficient stage-aware hardware adaptation with minimal runtime overhead.

\begin{figure*}[t]
\centering
\setlength{\abovecaptionskip}{4pt}
\setlength{\belowcaptionskip}{-4pt}
\setlength{\floatsep}{6pt}
\setlength{\textfloatsep}{6pt}
\begin{subfigure}{0.32\textwidth}
  \centering
  \includegraphics[width=\linewidth]{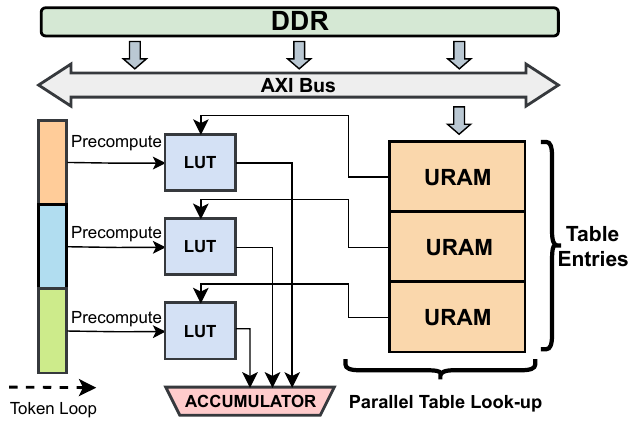}
  \caption{Table-look-up Matmul Engine}
  \label{fig:tlmm}
\end{subfigure}\hfill
\begin{subfigure}{0.35\textwidth}
  \centering
  \includegraphics[width=\linewidth]{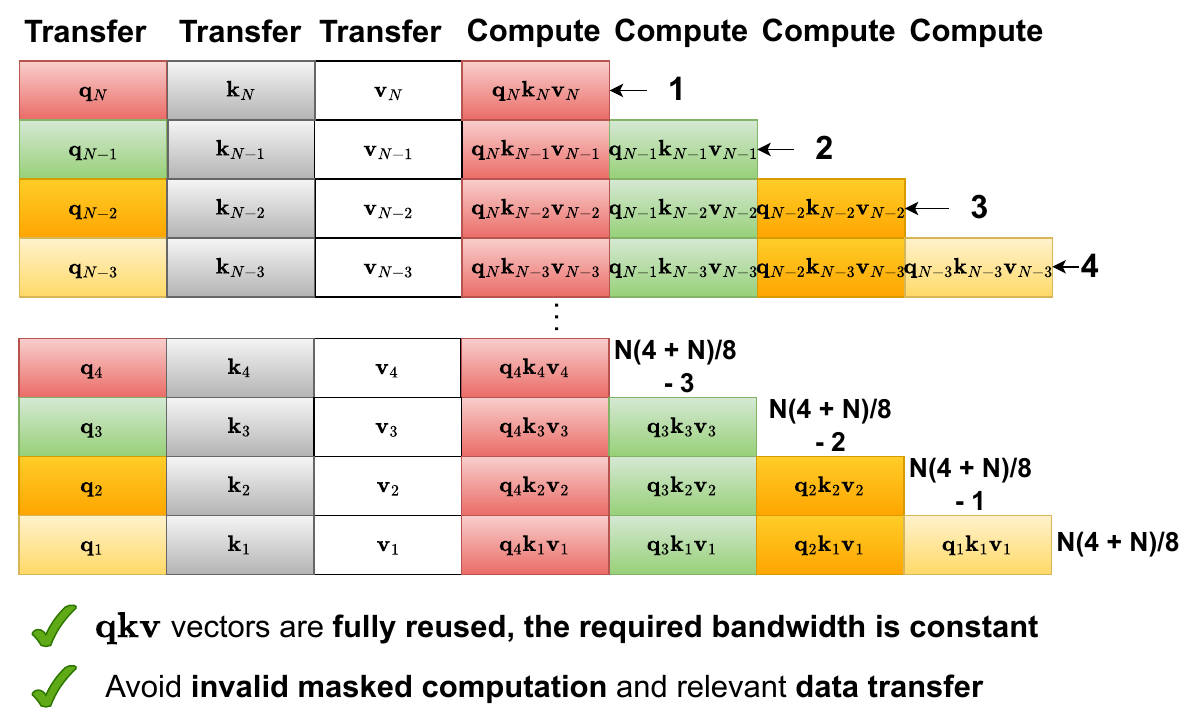}
  \caption{Reverse Attention Scheduling ($N_{\text{pe}}{=}4$)}
  \label{fig:reverse_schedule}
\end{subfigure}\hfill
\begin{subfigure}{0.32\textwidth}
  \centering
  \includegraphics[width=\linewidth]{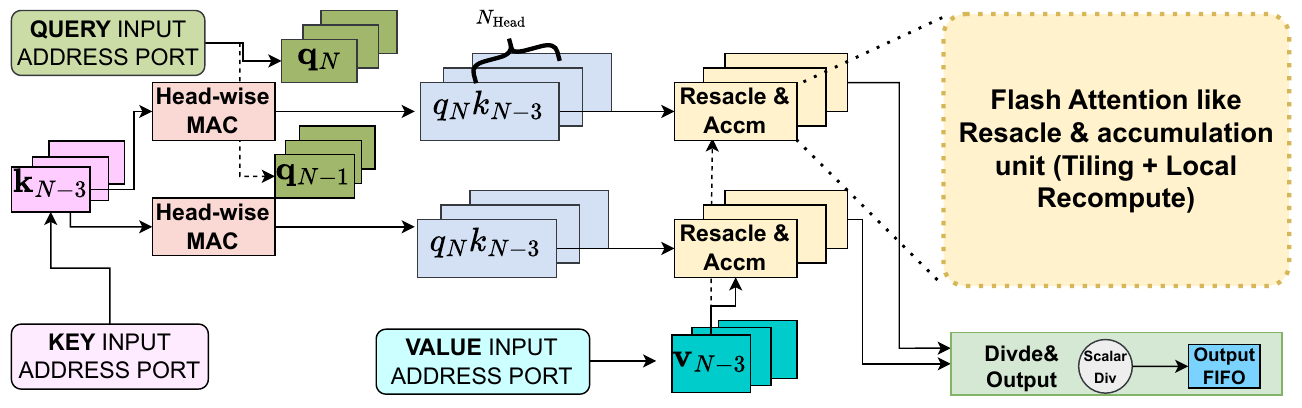}
  \caption{Prefill Attention Architecture ($N_{\text{pe}}{=}2$)}
  \vspace{1mm}
  \centering
  \includegraphics[width=\linewidth]{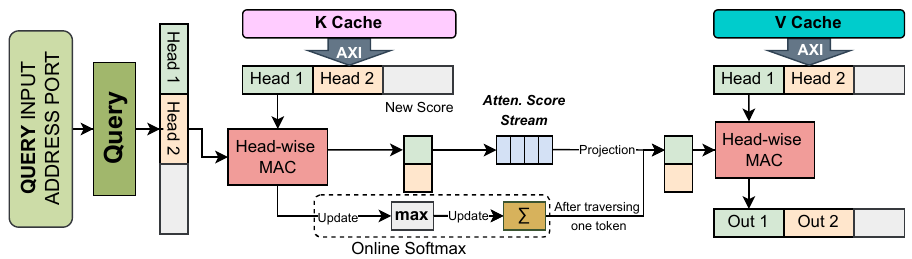}
  \caption{Decode Attention Architecture}
  \label{fig:decode_schedule}
\end{subfigure}
 \caption{Design and Optimization for Critical Modules in PD-Swap}
\label{fig:attention_triptych}
\end{figure*}

\subsubsection{Specialized Dataflow for Major Components in PD-Swap}

The prefill and decode stages share a unified data orchestration strategy for linear projections, utilizing the Table Lookup Linear mechanism (depicted in Figure \ref{fig:tlmm}). In this design, ternary weights are permanently resident on-chip, utilizing the high capacity of RAM. A distinct advantage of this architecture is that it eliminates the need for complex weight-reuse tiling schemes typically required to mitigate off-chip memory bandwidth bottlenecks. Since weights are already available on-chip, we do not need to block the computation to amortize DRAM access costs. Instead, we adopt a token-wise GEMV approach for both stages:
\textbf{Prefill Stage:} Handled as a batch of independent GEMV operations, where $\mathbf{X} \in \mathbb{R}^{d_{\text{in}} \times N_{\text{tokens}}}$. This avoids the complex tiling associated with traditional blocked GEMM.
\textbf{Decode Stage:} Inherently persists as a single-vector GEMV, where $\mathbf{X} \in \mathbb{R}^{d_{\text{in}} \times 1}$.
Specifically, for every value group, add/subtract combinations are pre-computed, followed by a parallel lookup operation where the URAM-resident weights serve as indices to retrieve values from the pre-computed lookup table. This flow eliminates the overhead of reloading weight channels from DDR, ensuring that both prefill and decode stages fully leverage the bandwidth benefits of all-on-chip linear weights. Following the linear transformation, the output streams through element-wise operations such as SiLU, RoPE, residual addition, and dequantization.

Contrastingly, the attention mechanism requires distinct dataflows for the prefill and decode stages due to the use of FP16 precision for Q, K, and V vectors, which often exceeds on-chip URAM/BRAM capacity. This necessitates a split architecture to address the conflicting compute-bound (prefill) and memory-bound (decode) characteristics.
We employ a Flash Attention-like~\cite{dao2022flashattentionfastmemoryefficientexact} operation to exploit token-level parallelism and spatial reuse. As depicted in Figure \ref{fig:reverse_schedule}, we utilize a reverse scheduling order to efficiently handle causal masking. The core objective here is to maximize the reuse of loaded $\mathbf{Q}$ vectors across multiple $\mathbf{K}$ and $\mathbf{V}$ blocks. The overall computation follows the standard attention formulation:
$
\text{Attention}(\mathbf{Q}, \mathbf{K}, \mathbf{V}) = \text{softmax}\left(\mathbf{Q} \mathbf{K}^T/\sqrt{d_k}\right)\mathbf{V}
$.
The Flash Attention method computes the Softmax by processing blocks iteratively. For each block $j$, the partial output $\mathbf{O}^{(j)}$ and accumulated maximum $\mathbf{m}^{(j)}$ and sum $\mathbf{l}^{(j)}$ are updated. The block-wise update for the Softmax output scale factor $\mathbf{D}_i$ and the output block $\mathbf{O}_i$ is given by:
\begin{equation}
\text{\scriptsize $
\begin{cases}
    \mathbf{m}^{(j)} = \max\left(\mathbf{m}^{(j-1)}, \text{rmax}(\mathbf{L}^{(j)})\right) \\
    \mathbf{l}^{(j)} = e^{\mathbf{m}^{(j-1)} - \mathbf{m}^{(j)}} \mathbf{l}^{(j-1)} + e^{\text{rmax}(\mathbf{L}^{(j)}) - \mathbf{m}^{(j)}} \text{rsum}\left(e^{\mathbf{L}^{(j)} - \text{rmax}(\mathbf{L}^{(j)})}\right) \\
    \mathbf{O}^{(j)} = \text{diag}\left( e^{\mathbf{m}^{(j-1)} - \mathbf{m}^{(j)}} \right) \mathbf{O}^{(j-1)} + e^{\text{rmax}(\mathbf{L}^{(j)}) - \mathbf{m}^{(j)}} \left(e^{\mathbf{L}^{(j)} - \text{rmax}(\mathbf{L}^{(j)})}\right) \mathbf{V}^{(j)}
\end{cases}
$}
\end{equation}
Where $\mathbf{L}^{(j)} = \mathbf{Q}^{(i)} (\mathbf{K}^{(j)})^T$ is the partial attention score block, $\mathbf{P}^{(j)}$ is the normalized probability block, and $\mathbf{m}, \mathbf{l}, \mathbf{O}$ are the accumulated running maximum, exponent sum, and final output, respectively. This blocked approach maximizes $\mathbf{Q}$-vector reuse by keeping $\mathbf{Q}^{(i)}$ resident while iterating through $\mathbf{K}^{(j)}$ and $\mathbf{V}^{(j)}$.

In the decoding phase (Figure \ref{fig:decode_schedule}), the sequence length is one ($L=1$), meaning there is no opportunity for Q-vector reuse. The operation degenerates into a memory-bound vector-matrix process against the accumulated Key-Value cache:
$
\mathbf{q}_{t} \cdot \mathbf{K}_{<t}^T \rightarrow \text{softmax} \rightarrow \cdot \mathbf{V}_{<t} \rightarrow \mathbf{o}_t
$
Here, a single query vector $\mathbf{q}_t \in \mathbb{R}^{1 \times d}$ must attend to the entire history of keys $\mathbf{K}_{<t} \in \mathbb{R}^{t \times d}$ and values $\mathbf{V}_{<t} \in \mathbb{R}^{t \times d}$ to produce the single output vector $\mathbf{o}_t$. Given the divergent hardware requirements—computation-heavy for prefill versus bandwidth-heavy for decode—we employ partial reconfiguration to dynamically swap the FPGA logic in the dynamic region, ensuring maximum utilization of LUT resources for each specific stage.

\subsubsection{Memory Port Utilization Optimization in Dynamic Region}
To further address the load asymmetry between prefill and decoding, we optimize the memory-interface allocation on the KV260 platform, which provides four High-Performance (HP) DDR ports. Our baseline design uses all four ports in both static and dynamic regions to stream weights and intermediate tensors. While this achieves good utilization during prefill, the decode stage exhibits severe imbalance: Q and output involve only a single token, whereas K and V require continuous access to the growing KV cache.

To address this, we redesign the HP-port mapping specifically for decode attention. Instead of assigning ports to QKV and output as in~\cite{qiao2025tellmev2}, we allocate two HP ports to K and two to V, doubling the available bandwidth for KV-cache reads. During kernel launch, the controller temporarily blocks other memory requests, bypasses one port to stream the Q token directly into on-chip buffers, and stores the output token locally before writing it back to DDR after KV transfers complete. This eliminates port contention and improves the effective decode stage bandwidth by nearly 2$\times$, ensuring that the memory subsystem matches the optimized compute structure of the decoding attention accelerator.

\begin{figure}[t]
\centering
\setlength{\abovecaptionskip}{4pt}
\setlength{\belowcaptionskip}{-4pt}
\setlength{\floatsep}{6pt}
\setlength{\textfloatsep}{6pt}
\begin{subfigure}{0.46\textwidth}
  \centering
  \includegraphics[width=\linewidth]{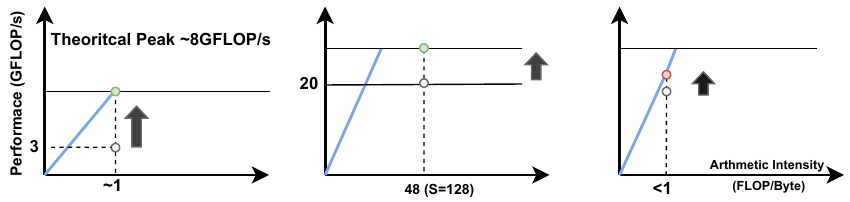}
  \caption{Qualitative Roofline Analysis for decoding attention (left), prefill attention (middle) and linear (right)}
  \label{fig:roofline}
\end{subfigure}\vspace{1mm}
\begin{subfigure}{0.45\textwidth}
  \centering
  \includegraphics[width=\linewidth]{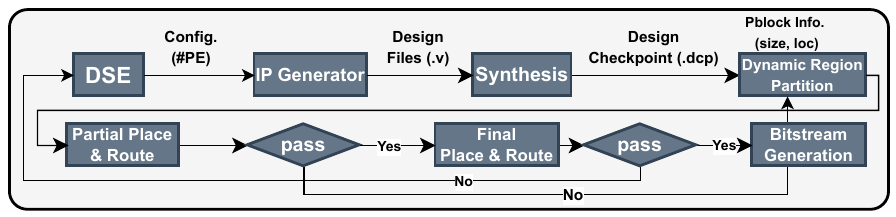}
  \caption{Automated Implemenetation Flow Chart}
  \label{fig:flowchart}
\end{subfigure}
 \caption{PD-Swap DSE and Automation Flow}
\label{fig:DSE}
\end{figure}

\subsection{Design Space Exploration and Automated Implementation Flow}
\subsubsection{Roofline Model Guided Design Choice}

The static–dynamic partition enabled by DPR significantly broadens the design space on resource-constrained FPGAs. By reclaiming idle logic and flexibly reallocating resources between the dynamic prefill/decoding attention modules and the static linear operators, designers can better align hardware capability with the computational characteristics of LLM inference. To guide these decisions, we construct qualitative roofline models for the major kernels, as shown in Figure~\ref{fig:roofline}. 

Prefill attention exhibits high arithmetic intensity and lies firmly in the compute-bound region. Decode attention, however, has extremely low arithmetic intensity and should ideally operate in the memory-bound regime, yet static designs lack the reusable resources needed to accelerate it effectively. In contrast, the decode-stage linear modules receive sufficient compute and bandwidth resources in our design and operate close to their roofline limits. Thus, decode attention becomes the dominant bottleneck, underscoring the importance of DPR-enabled resource redistribution.

\subsubsection{Resource Constraints and Objective Formulation}

We denote the resources allocated to the prefill projection, prefill attention, and decode attention modules as $\mathbf{r}_{proj}, \mathbf{r}_{atten}^{\text{pre}},  \mathbf{r}_{atten}^{\text{dec}}$
respectively. Since the prefill-attention and decode-attention accelerators time-share the same reconfigurable partition, their resource usage satisfies:
\begin{equation}
\mathbf{r}_{proj} + \max\{\mathbf{r}_{atten}^{\text{pre}}, \mathbf{r}_{atten}^{\text{dec}}\} \le \mathbf{R_{total}}.
\label{eq:resource_constraint}
\end{equation}

To normalize the impact of different configurations on execution time, we use a set of coefficients $P_{proj}$, $P_{atten}$, $D_{proj}$, and $D_{atten}$, to represent the projection and attention computing cost in prefill and decoding stage, which are empirically measured under a baseline hardware configuration.

For a sequence length $L$, the prefill stage consists of linearly scaling projection layers and a quadratically scaling attention kernel:
\begin{equation}
T_{\text{pre}} 
= \frac{P_{proj} L}{f_{pre}(\mathbf{r}_{proj})} 
+ \frac{P_{atten} L^{2}}{g_{pre}(\mathbf{r}_{atten}^{\text{pre}})}
+ T_{weights},
\label{eq:prefill_latency}
\end{equation}
where $f(\cdot)$ and $g(\cdot)$ are increasing functions of compute units, bandwidth, and hardware parallelism, and $T_{weights}$ is the inherent model weights loading time, which does not change with $L$. To maintain user-perceived responsiveness, we need to make sure:
\begin{equation}
T_{\text{pre}} \le T_{\text{pre}}^{\max}.
\label{eq:prefill_bound}
\end{equation}

The decode-stage latency can be formulated as follows:
\begin{equation}
T_{\text{dec}} 
= \frac{D_{proj}}{f_{dec}(\mathbf{r}_{proj})}
+ \frac{D_{atten} L}{g_{dec}(\mathbf{r}_{atten}^{\text{dec}})} 
+ T_{weights},
\label{eq:decode_latency}
\end{equation}
where $g_{dec}(\cdot)$ captures effective K/V bandwidth and compute parallelism.
We formulate the end-to-end latency minimization problem as:
\begin{equation}
\begin{aligned}
\min
\quad & T_{\text{pre}} + \alpha T_{\text{dec}}(L_{long}) + (1-\alpha) T_{\text{dec}}(L_{short})\\
\text{s.t.} \quad &
T_{\text{pre}} \le T_{\text{pre}}^{\max}, \\
& \mathbf{r}_p + \max\{\mathbf{r}_{a}^{\text{pre}}, \mathbf{r}_{a}^{\text{dec}}\} \le \mathbf{R}, \\
\end{aligned}
\label{eq:optimization}
\end{equation}
We use a weighted decoding latency ($\alpha = 0.7$) to ensure performance under long sequences. We profile each module across a wide range of configurations, and then collected results to perform the above design space exploration.


\subsubsection{Automated Hardware Implementation Flow}
To enable rapid iteration, we develop an automated build flow that connects DSE outputs to hardware generation, synthesis, and implementation. Our attention engine follows a spatial dataflow design, where PE count and parallelism configuration are exposed as tunable high-level parameters. After DSE selects the optimal parameters, these values are inserted into a parameterized hardware template to generate synthesizable HLS-based IP cores.

We then perform module-level synthesis for the linear projection, prefill attention, and decoding attention components. The dynamic region is defined as a dedicated pblock, whose size and placement have substantial influence on routing feasibility and timing. To validate feasibility, we first conduct standalone implementation of the dynamic-region module. Once timing and resource utilization are satisfied, we proceed with full-system place and route, integrating both the static region and the dynamic partition.

If overall timing closure still fails due to congestion or routing pressure, we iteratively reduce resource utilization in the dynamic partition---for example, by decreasing PE count or parallelism---and regenerate the hardware. This automated DSE-to-implementation flow ensures deployment of stage-specialized LLM hardware accelerators on FPGA fabrics, fully leveraging the flexibility enabled by dynamic partial reconfiguration.

\begin{table*}[!ht]
\centering
\caption{Unified cross-platform and FPGA-based comparison for edge LLM inference. WikiText-2 assess model performance. Resource utilization is reported for FPGA works.}
\vspace{-2mm}
\label{tab:unified-llm-comparison}
\setlength{\tabcolsep}{3pt}
\renewcommand{\arraystretch}{0.85}
\resizebox{\textwidth}{!}{%
\begin{tabular}{c c c c c c c c c c c c c c c c}
\toprule
\multirow{2}{*}{\textbf{Work}} &
\multirow{2}{*}{\textbf{Platform}} &
\multirow{2}{*}{\textbf{Processor}} &
\multirow{2}{*}{\textbf{Model}} &
\multirow{2}{*}{\textbf{Bitwidth}} &
\multicolumn{5}{c}{\textbf{FPGA Resource Utilization}} &
\multirow{2}{*}{\textbf{Power (W)}} &
\multirow{2}{*}{\textbf{WT-2 (PPL)}} &
\multicolumn{2}{c}{\textbf{Throughput (TK/S)}} &
\multicolumn{2}{c}{\textbf{Energy Efficiency (TK/J)}} \\
\cmidrule(lr){6-10} \cmidrule(lr){13-14} \cmidrule(lr){15-16}
& & & & & \textbf{LUT} & \textbf{FF} & \textbf{DSP} & \textbf{BRAM} & \textbf{URAM} & & & \textbf{Prefill} & \textbf{Decode} & \textbf{Prefill} & \textbf{Decode} \\
\midrule
Raspberry Pi 5 \cite{adafruit2025qwen3} & SoC & $4\times$ Cortex-A76 & Qwen 0.6B & W4-A16 &   — & — & — & — & — & 7.8 & 24.00 & 61.8 & 16.6 & 7.92 & 2.12 \\
Jetson Orin Nano \cite{jetsonailab2025slm} & GPU SoC & $8\times$ GPU SM & TinyLLaMA 1.1B & W4-A16 &  — & — & — & — & — & 25 & 12.42 & 324.9 & 67.6 & 12.9 & 2.70 \\
\midrule
LLaMAF \cite{LLaMAf} & FPGA SoC & ZCU102 & TinyLLaMA 1.1B & W8-A8 & 164K & 171K & 528 & 223 & — & 5.1 & 8.89 & — & 1.5 & — & 0.29 \\
MEADOW \cite{moitra2025meadow} & FPGA SoC & ZCU102 & OPT 1.3B & W8-A8 & 150K & — & 845 & 2034 & — & 10 & 15.41 & 100 & 2 & 10 & 0.20 \\
TeLLMe \cite{qiao2025tellmev2} & FPGA SoC & KV260 & BitNet 0.73B & W1.58-A8 & 98K & 137K & 610 & 98.5 & 60 & 4.8 & 12.79 & 143 & 25 & 29.8 & 5.2 \\
\midrule
\textbf{PD-Swap} & FPGA SoC & KV260 & BitNet 0.73B & W1.58-A8 & 102K & 176K & 750 & 124.5 & 62 & \textbf{4.9} & \textbf{12.79} & \textbf{148} & \textbf{27.8} & \textbf{30.2} & \textbf{5.67} \\
\bottomrule
\end{tabular}
}
\end{table*}

\subsection{Runtime Logic Swap Optimization}
To further reduce the runtime cost of hardware swapping, we introduce a latency-overlapped reconfiguration mechanism that exploits the execution structure of LLM inference. On AMD FPGAs, the reconfiguration time is directly proportional to the bitstream size. Although partial bitstreams reduce the latency to the millisecond range, prior work has shown that repeated reconfiguration within a single service request can still incur unacceptable overhead for latency-sensitive applications~\cite{quant_dpr}. In our LLM inference system, each request requires only a single reconfiguration (prefill attention \textrightarrow  decode attention), but multiple short-token requests in edge scenarios may still expose noticeable delays.

\begin{figure}
    \centering
    \includegraphics[width=1\linewidth]{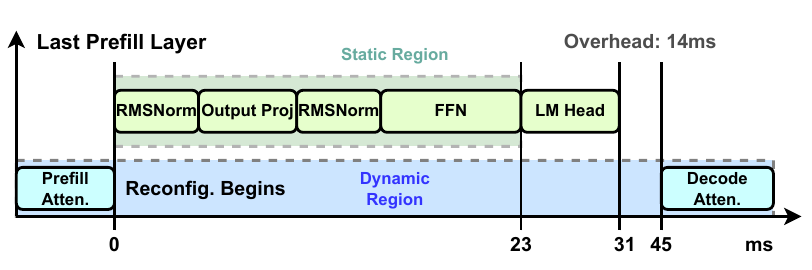}
    \caption{Latency-overlapped Rumtime Reconfiguration Mechanism (token length = 128)}\vspace{-1em}
    \label{fig:reconfig_overlap}
\end{figure}

To mitigate this, we leverage the intrinsic computation pipeline of transformer layers and overlap reconfiguration with the tail end of prefill computation. As shown in Figure~\ref{fig:reconfig_overlap}, the key observation is that prefill attention is no longer needed after the final layer’s attention computation, while subsequent operations (e.g., output projection and the entire FFN block) still incur substantial latency. We add a lightweight controller to the prefill attention module that, upon detecting the completion of the final attention layer, immediately signals the PS to initiate partial reconfiguration—without waiting for the remaining prefill operators to finish.

Measurements on our KV260 platform show that reconfiguring the attention module requires approximately 45 ms, whereas the remaining projection and FFN computations for a sequence length of 128 require roughly 31 ms. By initiating reconfiguration early, we successfully overlap most of the delay and reduce the effective reconfiguration overhead by about 75\%, making the cost of logic swapping nearly negligible. Although decoding begins with several non-attention operations that could, in principle, be further overlapped, we conservatively start the decoding stage only after confirming that the decode-attention bitstream has been fully loaded to ensure correctness under all conditions.

\section{Experimental Results and Analysis}
\subsection{Experimental Setup}
We evaluate the proposed system on the AMD Kria KV260 FPGA (Zynq UltraScale+ XCK26 MPSoC). The IP templates are implemented in High-level Synthesis (HLS) C/C++ and synthesized into RTL IPs using Vitis HLS 2024.1 tool. The runtime reconfigurable system is developed and implemented in Vivado 2024.1.
We use the BitNet 0.73B model~\cite{bitnet158} as our base LLM model. For fair comparison, we select prior works that implement approximately 1B-parameter models on edge platforms.
To evaluate the effectiveness of our runtime logic swapping architecture, we compare PD-Swap against TeLLMe~\cite{qiao2025tellmev2}, the state-of-the-art end-to-end LLM inference accelerator on edge FPGA, which deploys the same model with a static architecture on the same FPGA. 
\subsection{PD-Swap End-to-End Inference Performance}

\begin{figure}
    \centering
    \includegraphics[width=\linewidth]{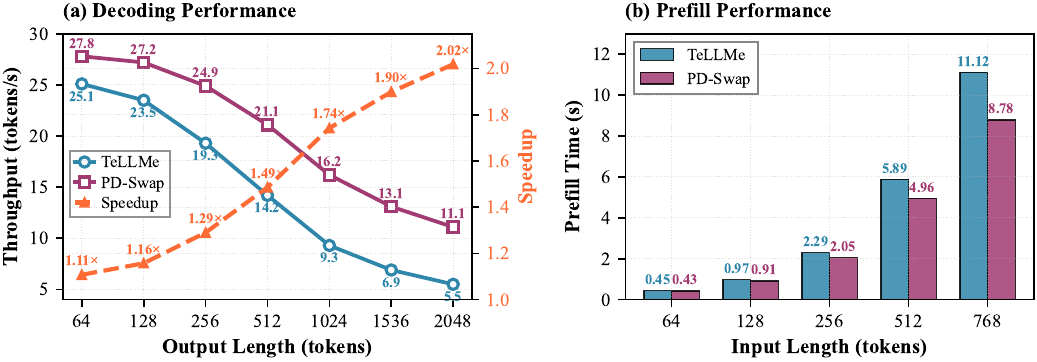}
    \caption{PD-Swap Inference Performance: (a) Decoding Throughput; (b) Prefill Time (Time-to-First-Token, TTFT).}\vspace{-2em} 
    \label{fig:throughput_plot}
\end{figure}
We first evaluate the LLM inference performance.
Figure~\ref{fig:throughput_plot} shows the LLM decoding (auto-regressive) throughput and the prefill time (Time-to-First-Token, TTFT) for both PD-Swap and TeLLMe. Our PD-Swap accelerator consistently outperforms the baseline across both decoding and prefill stages. In decoding, throughput improves from 1.11× at 64 tokens to 2.02× at 2048 tokens, showing increasing benefits as sequence length grows. Prefill latency is also reduced by 20–25\%, e.g., from 11.10 s to 8.80 s at 768 tokens. These results demonstrate that dynamically separating and reconfiguring hardware for prefill and decoding effectively alleviates performance bottlenecks and accelerates end-to-end LLM inference.

\subsection{Cross-Platform and FPGA-based Comparison for Edge LLM Inference}

As shown in Table \ref{tab:unified-llm-comparison}, our design demonstrates significant advantages over state-of-the-art baselines. 
In cross-platform comparisons, we outperform edge Platform Raspberry Pi 5. While the Jetson Orin Nano benefits from higher LPDDR5 bandwidth to achieve faster throughput, our FPGA solution demonstrates superior energy efficiency despite the bandwidth limitations of LPDDR4. Similarly, compared with LLaMAF, our framework achieves better overall performance. Furthermore, while MEADOW suffers from the inefficiencies of a unified architecture, our heterogeneous design addresses the disparity between prefill and decode. Compared to TeLLMe, our resource-efficient attention mechanism overcomes hardware bottlenecks, delivering higher performance, particularly in long-context scenarios.
Our design exhibits significantly higher performance in long-context scenarios compared to the most competitive baseline, TeLLMe. PD-Swap maintains >10 token/s decoding throughput even at 2048 input prompt length, while TeLLMe drop to ~5 token/s.

\begin{table}[t]
\centering
\caption{FPGA resource consumption breakdown.}
\label{tab:utilization}
\vspace{-2mm}
\setlength{\tabcolsep}{5pt}
\renewcommand{\arraystretch}{0.8}
\resizebox{\columnwidth}{!}{%
\begin{tabular}{lrrrrr}
\toprule
\textbf{Module} & \textbf{LUT} & \textbf{FF} & \textbf{BRAM} & \textbf{URAM} & \textbf{DSP} \\
\midrule
Table Lookup Linear Unit          & 42{,}854 & 50{,}752 &  5.5 &  0  & 320 \\
RMSNorm \& Find Max Unit&  6{,}210 & 11{,}206 &  4.0 &  4  &  47 \\
Other                   &  21{,}432   &  22{,}402  &34    &  48   &  5 \\
\midrule
\textbf{Dynamic Region} & \textbf{32{,}140} & \textbf{92{,}080} & \textbf{81} & \textbf{10} & \textbf{378} \\
\quad Prefill Attention        & 28{,}400 & 42{,}053 &140 &  8  & 303 \\
\quad Decoding Attention       & 26{,}418 & 27{,}236 & 16 &  8  & 278 \\
\midrule
\textbf{Total}                   & \textbf{102{,}102} & \textbf{176{,}440} & \textbf{124.5} & \textbf{62} & \textbf{750} \\
\textit{Utilization}             & \textit{(87\%)}   & \textit{(36\%)}     & \textit{(85\%)} & \textit{(96\%)} & \textit{(60\%)} \\
\textbf{Equivalent Total}    & \textbf{124{,}780} & \textbf{136{,}721} & \textbf{98.5} & \textbf{62} & \textbf{953} \\
\textit{Equivalent Utilization}             & \textit{(106\%)}   & \textit{(36\%)}     & \textit{(85\%)} & \textit{(96\%)} & \textit{(76\%)} \\
\bottomrule
\end{tabular}%
}
\vspace{-2mm}
\end{table}

As shown in Table~\ref{tab:utilization}, partial reconfiguration enables the Prefill and Decoding Attention modules to time-multiplex a unified dynamic area. This strategy avoids the need to shrink modules for simultaneous fit; instead, it allocates the maximum available resources to the active stage. The "Equivalent Total" of 106\% highlights this advantage, demonstrating that we can implement logic complexity exceeding the chip's static capacity, ensuring neither phase is performance-constrained by resource partitioning.

\section{Conclusion}

We propose a disaggregated ternary LLM accelerator that uses DPR to resolve prefill-decode asymmetry on edge FPGAs. By time-multiplexing specialized attention modules on the KV260, we alleviate LUT and URAM bottlenecks, enabling higher parallelism and longer context support. 
Our design achieves state-of-the-art performance, with up to 27~tokens/s decoding throughput. Compared with the prior static accelerators, PD-Swap achieves a 1.3$\times$--2.1$\times$ speedup, with larger gains at longer context lengths.
\bibliographystyle{unsrt}
\bibliography{references}

\appendix

\end{document}